# Electromagnetic Waves at a Charged and Lossy Planar Interface Determined by the Tangential Electric Field of Incident Plane Wave

Zhili Lin (林志立), *IEEE Senior Member*

***Abstract*—Based on the normal and tangential decomposition of both the wave vectors and the electric fields of plane electromagnetic waves at a charged and lossy planar interface, all of the incident, reflected, and refracted plane waves are found to be only determined by the tangential electric field of the incident plane wave. The complex wave vectors are easily calculated from the tangential wave vector of the incident plane wave based on the complex angles given by the complex Snell's law. The electric field magnitudes of the incident, reflected and refracted waves are directly derived from the tangential electric field magnitude of the incident plane based on the relationship of the two different types of decomposition and the continuous boundary condition of tangential electric fields. The time-averaged Poynting vectors and the surface Joule heat density at the interface are also given to demonstrate the validity of the formulation by the energy balance condition together with a specific example. This work directly proves the uniqueness theorem of time-varying electromagnetic fields and opens a new and fast route for calculating the reflected and transmitted waves at a charged and lossy planar interface without the need to perform the polarization decomposition of the incident plane wave.**

## I. Introduction

The reflection and refraction of electromagnetic waves at a planar interface between two different media are of fundamental importance in electromagnetics and optics [1, 2]. For example, many optical devices such as eyeglasses, contact lenses, and cameras are based on the characteristics of light waves undergoing reflection or refraction [3, 4]. The Snell's law and Fresnel equations are usually applied to investigate the reflection and refraction at a planar interface. Snell's law gives the intrinsic relationship between the angles of incidence and refraction with respect to the normal vector of the interface. The traditional practice for calculating the reflected and refracted waves from a given polarized incident wave is based on the polarization decomposition, where the incident plane wave is usually decomposed into two waves. One wave called the transverse electric (TE)-wave or also s-polarized wave has its electric field parallel to the interface and vertical to the plane of incidence. The other wave called the transverse magnetic (TM)-wave or also p-polarized wave has its electric field polarized in the plane of incidence and its magnetic field is parallel to the plane of the interface. Fresnel equations specify the amplitude coefficients for reflection and transmission at a perfectly flat and clean interface between two transparent or lossy homogeneous media for the two different polarizations. When the materials on one or both sides of the interface are lossy media with complex material parameters, the Snell's law and Fresnel equations should be written in complex form.

Manuscript received April 20, 2025. This work was supported in part by the Natural Science Foundation of Xiamen City of China under Grant 3502Z202473050 and by the National Natural Science Foundation of China under Grant 61101007.

The author is with the Fujian Key Laboratory of Light Propagation and Transformation, College of Information Science and Engineering, Huaqiao University, Xiamen 361021, China (e-mail: zllin2008@gmail.com).

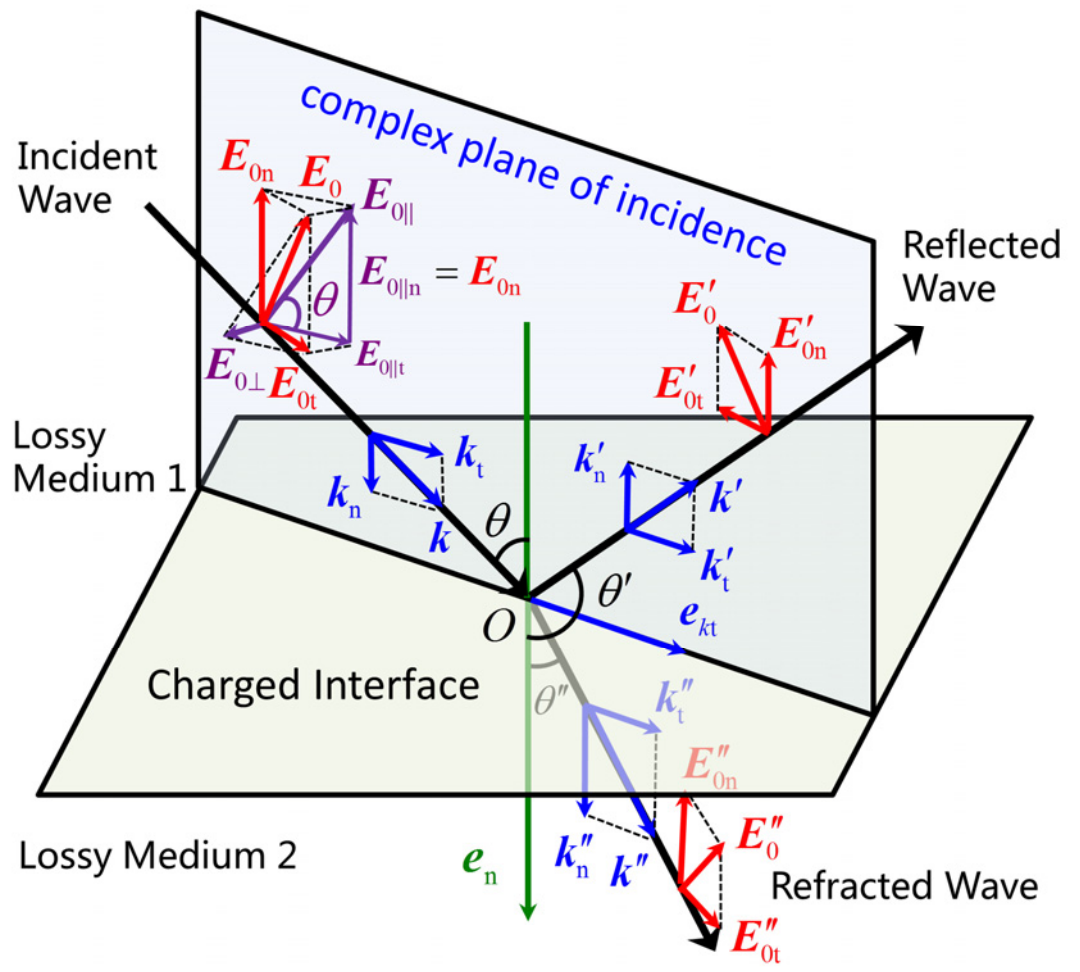

Fig. 1. Schematic of the reflection and refraction of a plane wave obliquely incident on a charged interface between two isotropic lossy media with the unit normal vector $\boldsymbol{e}_{\mathrm{n}}$. The physical positions of all the electric field vectors and their various components actually locate at the same reference point $O$.

In that case, the wave vectors, as well as the angles of incidence, reflection and refraction, are all with complex values [5, 6].

In this work, we try to decompose both the wave vectors and the electric fields of the incident, reflected, and refracted plane waves into the tangential and normal components with respect to the unit normal vector of the planar interface, which is different from the conventional polarization decomposition with respect to the plane of incidence. This allows the direct utilization of the continuous boundary condition of tangential electric fields and avoids the polarization decomposition of the incident plane wave. Thus a new and fast route for calculating the reflected and refracted waves from the given incident plane wave of arbitrarily polarization state at a charged and lossy planar interface is proposed.

## II. Decomposition of Complex Wave Vectors

As depicted in Fig.1, suppose that an arbitrary plane wave obliquely impinges upon a charged interface from the lossy medium 1 to the lossy medium 2 and the unit normal vector $\boldsymbol{e}_{\mathrm{n}}$ of interface is also pointing from medium 1 to medium 2. If the harmonic time-dependent factor $\exp(-\mathrm{j}\omega t)$ is suppressed, the complex electric fields of the incident, reflected and refracted plane waves propagating in the two isotropic lossy media are expressed as

$$\boldsymbol{E}(\boldsymbol{r}) = \boldsymbol{E}_0\, \mathrm{e}^{\mathrm{j}\boldsymbol{k}\cdot(\boldsymbol{r}-\boldsymbol{r}_0)},\ \boldsymbol{E}'(\boldsymbol{r}) = \boldsymbol{E}'_0\, \mathrm{e}^{\mathrm{j}\boldsymbol{k}'\cdot(\boldsymbol{r}-\boldsymbol{r}_0)},\ \boldsymbol{E}''(\boldsymbol{r}) = \boldsymbol{E}''_0\, \mathrm{e}^{\mathrm{j}\boldsymbol{k}''\cdot(\boldsymbol{r}-\boldsymbol{r}_0)} \quad (1)$$

respectively, where $\boldsymbol{r}_0$ is the position vector of the reference point $O$ on the interface and $\boldsymbol{E}_0 = \boldsymbol{E}(\boldsymbol{r}_0)$, $\boldsymbol{E}'_0 = \boldsymbol{E}'(\boldsymbol{r}_0)$ and $\boldsymbol{E}''_0 = \boldsymbol{E}''(\boldsymbol{r}_0)$ are the complex electric field magnitudes at point $O$. $\boldsymbol{k}$, $\boldsymbol{k}'$ and $\boldsymbol{k}''$ are the wave vectors of the incident, reflected and transmitted waves,

$$\boldsymbol{k} = k\boldsymbol{e}_k,\ \boldsymbol{k}' = k'\boldsymbol{e}'_k,\ \boldsymbol{k}'' = k''\boldsymbol{e}''_k \quad (2)$$

where $k = k' = k_1 = \omega\sqrt{\mu_1\varepsilon_1}$ and $k'' = k_2 = \omega\sqrt{\mu_2\varepsilon_2}$ are the corresponding wave numbers, respectively. For the two lossy media, the complex effective permittivities and the complex permeabilities are given by

$$\varepsilon_{1,2}(\omega) = \varepsilon'_{1,2}(\omega) + \mathrm{j}\varepsilon''_{1,2}(\omega) + \mathrm{j}\sigma_{1,2} / \omega \tag{3}$$

$$\mu_{1,2}(\omega) = \mu'_{1,2}(\omega) + \mathrm{j}\mu''_{1,2}(\omega) \tag{4}$$

respectively, where $\varepsilon'_{1,2}$ and $\varepsilon''_{1,2}$ are the real and imaginary parts of complex dielectric constants, $\sigma_{1,2}$ are the electrical conductivities, $\mu'_{1,2}$ and $\mu''_{1,2}$ are the real and imaginary parts of the complex permeabilities. For a homogeneous plane wave, $\boldsymbol{e}_k$ is a real-valued unit vector of $\boldsymbol{k}$ with a physically meaningful direction of wave propagation. However, for an inhomogeneous plane wave, $\boldsymbol{e}_k$ is a complex-valued unit vector without a physically meaningful direction and $\boldsymbol{k}$ is often represented as the superposition of the phase vector $\boldsymbol{\beta}$ and the attenuation vector $\boldsymbol{\alpha}$, $\boldsymbol{k} = \boldsymbol{\beta} + \mathrm{j}\boldsymbol{\alpha}$. Especially, when $\boldsymbol{\beta}$, $\boldsymbol{\alpha}$ and the unit normal vector $\boldsymbol{e}_\mathrm{n}$ of the interface are not coplanar, the complex plane of incidence includes two real planes, the real plane of $\boldsymbol{\beta}$ and the real plane of $\boldsymbol{\alpha}$ [7, 8].

As shown in Fig. 1, the complex wave vectors of the incident, reflected and refracted waves, $\boldsymbol{k}$, $\boldsymbol{k}'$ and $\boldsymbol{k}''$, are decomposed with respect to the unit normal vector $\boldsymbol{e}_\mathrm{n}$ of the planar interface into the normal components, $\boldsymbol{k}_\mathrm{n}$, $\boldsymbol{k}'_\mathrm{n}$ and $\boldsymbol{k}''_\mathrm{n}$, and the tangential components, $\boldsymbol{k}_\mathrm{t}$, $\boldsymbol{k}'_\mathrm{t}$ and $\boldsymbol{k}''_\mathrm{t}$, respectively. For example, $\boldsymbol{k}$ can be decomposed into the form

$$\boldsymbol{k} = \boldsymbol{k}_\mathrm{t} + (\boldsymbol{e}_\mathrm{n} \cdot \boldsymbol{k})\boldsymbol{e}_\mathrm{n} = \boldsymbol{k}_\mathrm{t} + \boldsymbol{k}_\mathrm{n} \tag{5}$$

based on the vector identity, where $\boldsymbol{k}_\mathrm{n} = (\boldsymbol{e}_\mathrm{n} \cdot \boldsymbol{k})\boldsymbol{e}_\mathrm{n} = k_\mathrm{n}\boldsymbol{e}_\mathrm{n}$ and $\boldsymbol{k}_\mathrm{t} = \boldsymbol{k} - \boldsymbol{k}_\mathrm{n} = k_\mathrm{t}\boldsymbol{e}_{kt}$ with the unit vector $\boldsymbol{e}_{kt}$ satisfying $\boldsymbol{e}_{kt} \cdot \boldsymbol{e}_{kt} = 1$. It is noted that for an inhomogeneous incident plane wave, $\boldsymbol{e}_{kt}$ is a complex-valued unit vector without a physically meaningful direction like that of $\boldsymbol{e}_k$. Also because of $\boldsymbol{e}_\mathrm{n} \cdot \boldsymbol{e}_{kt} = 0$, we have

$$\boldsymbol{k} \cdot \boldsymbol{k} = (\boldsymbol{k}_\mathrm{t} + \boldsymbol{k}_\mathrm{n}) \cdot (\boldsymbol{k}_\mathrm{t} + \boldsymbol{k}_\mathrm{n}) = k_\mathrm{t}^2 + k_\mathrm{n}^2 = k^2 = k_1^2 \tag{6}$$

where $k_\mathrm{t} = \sqrt{\boldsymbol{k}_\mathrm{t} \cdot \boldsymbol{k}_\mathrm{t}}$ and $k_\mathrm{n} = \sqrt{k_1^2 - k_\mathrm{t}^2}$ are the normal and tangential wave numbers of the incident wave, respectively. Since the trigonometric identity $\cos^2\theta + \sin^2\theta = 1$ holds even for a complex $\theta$, we define the complex angle of incidence with respect to $\boldsymbol{e}_\mathrm{n}$ as

$$\theta = \arcsin(k_\mathrm{t} / k_1) \tag{7}$$

with $k_1 = \omega\sqrt{\mu_1\varepsilon_1}$, and we have $k_\mathrm{n} = k\cos\theta$ and $k_\mathrm{t} = k\sin\theta$ based on (6). Thus the wave vector of incident wave $\boldsymbol{k}$ is related to its tangential component $\boldsymbol{k}_\mathrm{t}$ by

$$\boldsymbol{k} = \boldsymbol{k}_\mathrm{t} + \boldsymbol{k}_\mathrm{n} = \boldsymbol{k}_\mathrm{t} + k_1\cos\theta\boldsymbol{e}_\mathrm{n} \tag{8}$$

According to the phase matching condition at the interface of two lossy media, it can be derived that

$$\boldsymbol{k}_\mathrm{t} = \boldsymbol{k}'_\mathrm{t} = \boldsymbol{k}''_\mathrm{t} = \boldsymbol{k} - (\boldsymbol{e}_\mathrm{n} \cdot \boldsymbol{k})\boldsymbol{e}_\mathrm{n} = k_\mathrm{t}\boldsymbol{e}_{kt} \tag{9}$$

This yields the complex form of Snell's law given by

$$k_1\sin\theta = k_1\sin\theta' = k_2\sin\theta'' = k_\mathrm{t} \tag{10}$$

where $\theta'$ and $\theta''$ are the (possibly) complex angles of reflection and refraction defined by

$$\theta' = \pi - \theta\,,\ \theta'' = \arcsin(k_\mathrm{t} / k_2) \tag{11}$$

with $k_2 = \omega\sqrt{\mu_2\varepsilon_2}$. Then $k'_\mathrm{n} = k_1\cos\theta' = -k_1\cos\theta$, and the complex wave vector of the reflected wave is given by

$$\boldsymbol{k}' = \boldsymbol{k}'_\mathrm{t} + \boldsymbol{k}'_\mathrm{n} = \boldsymbol{k}_\mathrm{t} + k'_\mathrm{n}\boldsymbol{e}_\mathrm{n} = \boldsymbol{k}_\mathrm{t} - k_1\cos\theta\boldsymbol{e}_\mathrm{n} \tag{12}$$

Meanwhile, based on the complex angle $\theta''$ calculated by (11), we have $k''_\mathrm{n} = k_2\cos\theta''$, so that the complex wave vector of the refracted wave is obtained by

$$\boldsymbol{k}'' = \boldsymbol{k}''_\mathrm{t} + \boldsymbol{k}''_\mathrm{n} = \boldsymbol{k}_\mathrm{t} + k''_\mathrm{n}\boldsymbol{e}_\mathrm{n} = \boldsymbol{k}_\mathrm{t} + k_2\cos\theta''\boldsymbol{e}_\mathrm{n} \tag{13}$$

Therefore, based on (6)-(13), the complex wave vectors of the incident, reflected and refracted waves, $\boldsymbol{k}$, $\boldsymbol{k}'$ and $\boldsymbol{k}''$, are all determined by the tangential wave vector $\boldsymbol{k}_\mathrm{t}$ and its magnitude $k_\mathrm{t} = \sqrt{\boldsymbol{k}_\mathrm{t} \cdot \boldsymbol{k}_\mathrm{t}}$ of the incident plane wave.

## III. Decomposition of Electric Field Magnitudes

At the reference point $O$ on the interface, the electric field magnitude of the incident plane wave $\boldsymbol{E}_0$ is usually decomposed into the polarization form,

$$\boldsymbol{E}_0 = \boldsymbol{E}_{0\perp} + \boldsymbol{E}_{0\parallel} = E_{0\perp}\boldsymbol{e}_\perp + E_{0\parallel}\boldsymbol{e}_\parallel \tag{14}$$

where $E_{0\perp}$ is the vertical component of s polarization perpendicular to the complex incident plane and parallel to the interface, $E_{0\parallel}$ is the parallel component of p polarization parallel to the complex plane of incidence and perpendicular to the complex wave vector $\boldsymbol{k}$. On the other hand, $\boldsymbol{E}_0$ can be decomposed with respect to the unit normal vector $\boldsymbol{e}_\mathrm{n}$ into the form given by

$$\boldsymbol{E}_0 = \boldsymbol{E}_{0\mathrm{n}} + \boldsymbol{E}_{0\mathrm{t}} \tag{15}$$

where $\boldsymbol{E}_{0\mathrm{n}}$ is the vector of normal component and $\boldsymbol{E}_{0\mathrm{t}}$ is the vector of tangential component, respectively. According to Fig. 2, the tangential electric field magnitude can be written as

$$\boldsymbol{E}_{0\mathrm{t}} = \boldsymbol{E}_{0\perp} + \boldsymbol{E}_{0\parallel\mathrm{t}} = E_{0\perp}\boldsymbol{e}_\perp + E_{0\parallel}\cos\theta\boldsymbol{e}_{kt} \tag{16}$$

Similarly, we have the tangential electric field of the reflected wave,

$$\begin{aligned}\boldsymbol{E}'_{0\mathrm{t}} &= \boldsymbol{E}'_{0\perp} + \boldsymbol{E}'_{0\parallel\mathrm{t}} = E'_{0\perp}\boldsymbol{e}_\perp + E'_{0\parallel}\cos\theta'\boldsymbol{e}_{kt} \\ &= r_\perp E_{0\perp}\boldsymbol{e}_\perp - r_\parallel E_{0\parallel}\cos\theta\boldsymbol{e}_{kt}\end{aligned} \tag{17}$$

where $r_\perp$ and $r_\parallel$ are the Fresnel reflection coefficients for the s and p polarizations at a charged interface given by [9]

$$\begin{cases} r_\perp = \dfrac{E'_{0\perp}}{E_{0\perp}} = \dfrac{Z_2\cos\theta - Z_1\cos\theta'' - \sigma_\mathrm{s}Z_1Z_2}{Z_2\cos\theta + Z_1\cos\theta'' + \sigma_\mathrm{s}Z_1Z_2} \\ r_\parallel = \dfrac{E'_{0\parallel}}{E_{0\parallel}} = \dfrac{Z_1\cos\theta - Z_2\cos\theta'' + \sigma_\mathrm{s}Z_1Z_2\cos\theta\cos\theta''}{Z_1\cos\theta + Z_2\cos\theta'' + \sigma_\mathrm{s}Z_1Z_2\cos\theta\cos\theta''} \end{cases} \tag{18}$$

respectively. Here $Z_1 = \sqrt{\mu_1 / \varepsilon_1}$ and $Z_2 = \sqrt{\mu_2 / \varepsilon_2}$ are the

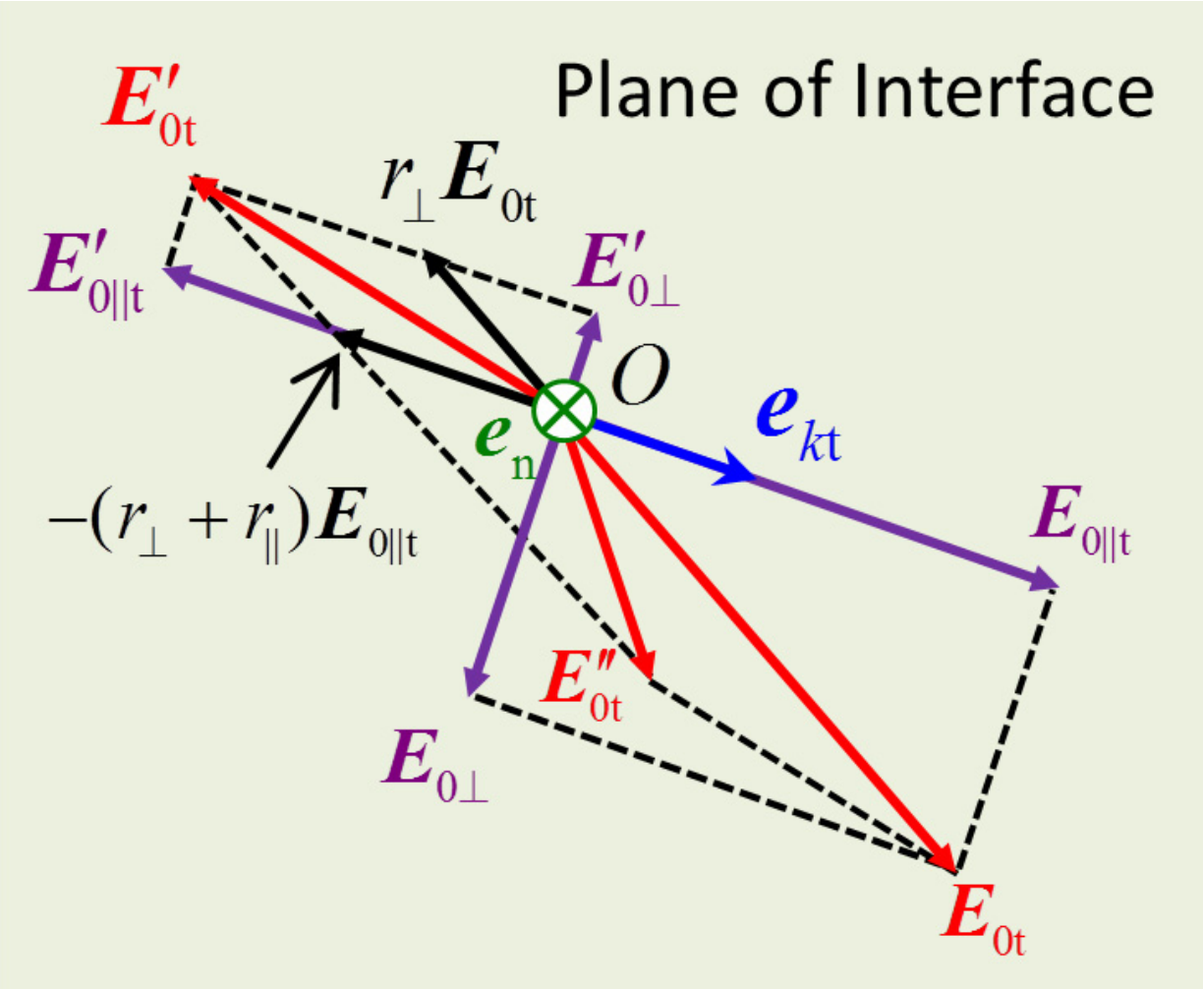

Fig. 2. The geometric relationship between the various tangential electric field components in the plane of interface reveals the continuous boundary condition of tangential electric fields at the interface of two different media.

complex intrinsic impedances of the two lossy media and $\sigma_s$ is the surface conductivity of interface proportional to the external surface charge density $\rho_s$ given by [10, 11]

$$\sigma_s(\omega) = \frac{\rho_s q_s}{m_s(\gamma_s - \mathrm{j}\omega)} \tag{19}$$

with $\gamma_s = k_B T / \hbar$, where $\rho_s$ is the external surface charge density, $q_s$ is the electric charge, $m_s$ is the mass of charge, $k_B$ is the Boltzmann constant, $T$ is the temperature in Kelvin, and $\hbar$ is the reduced Planck constant.

On multiplying (16) by $r_\perp$, and substituting the result about the item $r_\perp E_{0\perp} \boldsymbol{e}_\perp$ into (17), we obtain

$$\boldsymbol{E}'_{0t} = r_\perp \boldsymbol{E}_{0t} - (r_\perp + r_\|) E_{0\|} \cos\theta \boldsymbol{e}_{kt} \tag{20}$$

Since $\boldsymbol{e}_{kt} \cdot \boldsymbol{e}_\perp = 0$, the scalar product of $\boldsymbol{e}_{kt}$ and (16) gives

$$\boldsymbol{e}_{kt} \cdot \boldsymbol{E}_{0t} = E_{0\|} \cos\theta = E_{0\|t} \tag{21}$$

Then the substitution of (21) into (20) yields

$$\boldsymbol{E}'_{0t} = r_\perp \boldsymbol{E}_{0t} - (r_\perp + r_\|) E_{0\|t} \boldsymbol{e}_{kt} \tag{22}$$

where $\boldsymbol{e}_{kt} = \boldsymbol{k}_t / k_t = \boldsymbol{k}_t / \sqrt{\boldsymbol{k}_t \cdot \boldsymbol{k}_t}$ with $\boldsymbol{k}_t = \boldsymbol{k} - (\boldsymbol{e}_n \cdot \boldsymbol{k})\boldsymbol{e}_n$ or $\boldsymbol{k}_t = (\boldsymbol{e}_n \times \boldsymbol{k}) \times \boldsymbol{e}_n$. Eq. (22) is the most significant contribution of this work that reveals the relation between $\boldsymbol{E}'_{0t}$ and $\boldsymbol{E}_{0t}$. Based on the boundary condition of tangential electric fields, the tangential electric field magnitude of the refracted wave is easily obtained,

$$\boldsymbol{E}''_{0t} = \boldsymbol{E}_{0t} + \boldsymbol{E}'_{0t} \tag{23}$$

Moreover, according to Fig. 1 and based on (22) and (23), the normal electric field magnitudes of the incident, reflected and refracted waves are given by

$$\boldsymbol{E}_{0n} = \boldsymbol{E}_{0\|n} = -E_{0\|t} \tan\theta \boldsymbol{e}_n = -\tan\theta E_{0\|t} \boldsymbol{e}_n \tag{24}$$

$$\boldsymbol{E}'_{0n} = -\tan\theta'(\boldsymbol{e}_{kt} \cdot \boldsymbol{E}'_{0t})\boldsymbol{e}_n = -\tan\theta r_\| E_{0\|t} \boldsymbol{e}_n \tag{25}$$

$$\boldsymbol{E}''_{0n} = -\tan\theta''(\boldsymbol{e}_{kt} \cdot \boldsymbol{E}''_{0t})\boldsymbol{e}_n = -\tan\theta''(1 - r_\|) E_{0\|t} \boldsymbol{e}_n \tag{26}$$

with $E_{0\|t} = \boldsymbol{e}_{kt} \cdot \boldsymbol{E}_{0t}$, respectively. Therefore the total electric field magnitudes of the incident, reflected and refracted waves are finally acquired by the component combinations,

$$\boldsymbol{E}_0 = \boldsymbol{E}_{0n} + \boldsymbol{E}_{0t}, \; \boldsymbol{E}'_0 = \boldsymbol{E}'_{0n} + \boldsymbol{E}'_{0t}, \; \boldsymbol{E}''_0 = \boldsymbol{E}''_{0n} + \boldsymbol{E}''_{0t} \tag{27}$$

which are all determined by the tangential electric field magnitude $\boldsymbol{E}_{0t}$ and the unit direction vector $\boldsymbol{e}_{kt}$ of $\boldsymbol{k}_t$. It is worth noting that these formulas don't involve the Fresnel transmission coefficients.

It should be emphasized that the above formulas are derived based on the quantities at the interface. In fact, assuming that $\boldsymbol{r}_s$ is any point on the interface, the tangential electric field of the incident plane wave at the interface is given by

$$\boldsymbol{E}_t(\boldsymbol{r}_s) = \boldsymbol{E}_{0t} \, \mathrm{e}^{\mathrm{j}\boldsymbol{k}\cdot(\boldsymbol{r}_s - \boldsymbol{r}_0)} \tag{28}$$

It is noted that $\boldsymbol{e}_n \cdot (\boldsymbol{r}_s - \boldsymbol{r}_0) = 0$ as the difference vector $\boldsymbol{r}_s - \boldsymbol{r}_0$ is in the plane of interface. Meanwhile since $\boldsymbol{k} = \boldsymbol{k}_t + k_n \boldsymbol{e}_n$, we have $\boldsymbol{k} \cdot (\boldsymbol{r}_s - \boldsymbol{r}_0) = \boldsymbol{k}_t \cdot (\boldsymbol{r}_s - \boldsymbol{r}_0)$, so that

$$\boldsymbol{E}_t(\boldsymbol{r}_s) = \boldsymbol{E}_{0t} \, \mathrm{e}^{\mathrm{j}\boldsymbol{k}_t\cdot(\boldsymbol{r}_s - \boldsymbol{r}_0)} \tag{29}$$

Thus both the tangential electric field magnitude $\boldsymbol{E}_{0t}$ and the tangential wave vector $\boldsymbol{k}_t$ are included in the tangential electric field $\boldsymbol{E}_t$ of the incident wave. Therefore, the electric fields of the incident, reflected and transmitted waves, $\boldsymbol{E}$, $\boldsymbol{E}'$ and $\boldsymbol{E}''$, are all determined by the tangential electric field $\boldsymbol{E}_t$ at the interface.

In practice, if the incident plane wave is given with the wave vector $\boldsymbol{k}$ and the electric field magnitude $\boldsymbol{E}_0$, we can calculate the tangential wave vector by $\boldsymbol{k}_t = \boldsymbol{k} - (\boldsymbol{e}_n \cdot \boldsymbol{k})\boldsymbol{e}_n$ and the tangential electric field magnitude by $\boldsymbol{E}_{0t} = \boldsymbol{E}_0 - (\boldsymbol{e}_n \cdot \boldsymbol{E}_0)\boldsymbol{e}_n$. Then the electric fields of the reflected and refracted waves are obtained by the proposed methodology. Thereafter, the magnetic fields of the incident, reflected and refracted waves can be calculated by

$$\boldsymbol{H} = \frac{\boldsymbol{k} \times \boldsymbol{E}}{\omega\mu_1}, \; \boldsymbol{H}' = \frac{\boldsymbol{k}' \times \boldsymbol{E}'}{\omega\mu_1}, \; \boldsymbol{H}'' = \frac{\boldsymbol{k}'' \times \boldsymbol{E}''}{\omega\mu_2} \tag{30}$$

according to the Faraday's law of electromagnetic induction based on the previously obtained electric fields, respectively.

## IV. Energy Balance and Example

The validity and correctness of the above formulation can be verified by the energy balance condition derived from the complex Poynting theorem by applying a small Gaussian pillbox surrounding the charged interface given by [12]

$$\boldsymbol{e}_n \cdot \boldsymbol{S}^{\mathrm{M1}}_{\mathrm{av}} = \boldsymbol{e}_n \cdot \boldsymbol{S}^{\mathrm{M2}}_{\mathrm{av}} + p_s \tag{31}$$

Here $\boldsymbol{S}^{\mathrm{M1}}_{\mathrm{av}}$ is the time-averaged Poynting vector in medium 1 given by

$$\boldsymbol{S}^{\mathrm{M1}}_{\mathrm{av}} = \boldsymbol{S}_{\mathrm{av}} + \boldsymbol{S}'_{\mathrm{av}} + \boldsymbol{S}^{\mathrm{mix}}_{\mathrm{av}} = \frac{1}{2}\mathrm{Re}[(\boldsymbol{E} + \boldsymbol{E}') \times (\boldsymbol{H} + \boldsymbol{H}')^*] \tag{32}$$

where $\boldsymbol{S}_{\mathrm{av}}$ and $\boldsymbol{S}'_{\mathrm{av}}$ are the time-averaged Poynting vectors of the incident and reflected waves,

$$\boldsymbol{S}_{\text{av}} = \frac{1}{2}\text{Re}[\boldsymbol{E}\times\boldsymbol{H}^*],\ \boldsymbol{S}'_{\text{av}} = \frac{1}{2}\text{Re}[\boldsymbol{E}'\times\boldsymbol{H}'^*] \tag{33}$$

respectively, and $\boldsymbol{S}_{\text{av}}^{\text{mix}}$ is the mixed Poynting vector in the interference region of the incident and reflected waves,

$$\boldsymbol{S}_{\text{av}}^{\text{mix}} = \frac{1}{2}\text{Re}[\boldsymbol{E}\times\boldsymbol{H}'^* + \boldsymbol{E}'\times\boldsymbol{H}^*] \tag{34}$$

In medium 2, there only exists the refracted wave, so that the time-averaged Poynting vector is given by

$$\boldsymbol{S}_{\text{av}}^{\text{M2}} = \boldsymbol{S}''_{\text{av}} = \frac{1}{2}\text{Re}[\boldsymbol{E}''\times\boldsymbol{H}''^*] \tag{35}$$

The item $p_{\text{s}}$ is the surface Joule heat density at the lossy interface contributed by the surface current given by

$$p_{\text{s}} = \frac{1}{2}\text{Re}[\boldsymbol{J}_{\text{s}}\cdot\boldsymbol{E}_{\text{t}}^*] = \frac{1}{2}\text{Re}[\sigma_{\text{s}}\boldsymbol{E}_{\tan}\cdot\boldsymbol{E}_{\tan}^*] \tag{36}$$

where $\boldsymbol{E}_{\tan} = \boldsymbol{E}_{\text{t}} + \boldsymbol{E}'_{\text{t}} = \boldsymbol{E}''_{\text{t}}$ is the tangential electric field at the interface and $\sigma_{\text{s}}$ is the surface conductivity.

Finally, a specific example is presented to verify our proposed methodology and as well as to show the calculation procedure. As depicted in Fig. 3, a plane wave with frequency $f = 1$ GHz propagating in the lossy medium 1 is obliquely incident on a charged interface between two isotropic lossy media with the unit normal vector $\boldsymbol{e}_{\text{n}} = \boldsymbol{e}_z$. Suppose that the interface is charged by external electrons with a surface charge density $\rho_{\text{s}} = -2\times10^{-5}$ C/m$^2$, which is less than the surface charge density $\rho_{\text{s}} = 2.66\times10^{-5}$ C/m$^2$ corresponding to the air breakdown field strength $E_{\text{br}} = 3\times10^6$ V/m. Then according to the model given by (19), the surface conductivity is $\sigma_{\text{s}} \approx 1.04\times10^{-7}$+j$1.95\times10^{-11}$ S at frequency $f$. Meanwhile, the electromagnetic parameters of the two lossy media at frequency $f$ are arbitrarily assigned as $\varepsilon'_{\text{r1}} = 1.69$, $\varepsilon''_{\text{r1}} = 0.2$, $\sigma_1 = 0.3$ S/m, $\mu'_{\text{r1}} = 1.5$, $\mu''_{\text{r1}} = 0.5$ and $\varepsilon'_{\text{r2}} = 2.25$, $\varepsilon''_{\text{r2}} = 0.3$, $\sigma_2 = 0.5$ S/m, $\mu'_{\text{r2}} = 1$, $\mu''_{\text{r2}} = 0.2$, respectively.

Assume that the Cartesian coordinates are established on the reference point $O$ with the directions of the $x$, $y$, $z$ axes depicted in Fig. 3. The tangential electric field $\boldsymbol{E}_{\text{t}}$ at the interface is given by (29) with $\boldsymbol{r}_0 = 0$ and $\boldsymbol{r}_{\text{s}} = x\boldsymbol{e}_x + y\boldsymbol{e}_y$, where the tangential electric field magnitude $\boldsymbol{E}_{0\text{t}}$ is arbitrarily assumed that

$$\boldsymbol{E}_{0\text{t}} = E_{0\text{t}}(\cos\varphi_{Et}\boldsymbol{e}_x + \cos\varphi_{Et}\boldsymbol{e}_y) \tag{37}$$

with $E_{0\text{t}} = 100\text{e}^{\text{j}\pi/3}$ V/m and $\varphi_{Et} = 45^\circ$, and the tangential wave vector is arbitrarily assumed that

$$\boldsymbol{k}_{\text{t}} = k_{\text{t}}(\cos\varphi_{kt}\boldsymbol{e}_x + \sin\varphi_{kt}\boldsymbol{e}_y) \tag{38}$$

with $k_{\text{t}} = k_1/\sqrt{2}$ and $\varphi_{kt} = 30^\circ$ for an elliptically polarized homogenous incident wave. Then we get the real or complex angles of incidence, reflection and refraction based on (7) and (11) that

$\theta = 45^\circ$, $\theta' = 135^\circ$, $\theta'' = 0.758$ +j0.0325 rad where the numerical values are retained with 3 significant digits.

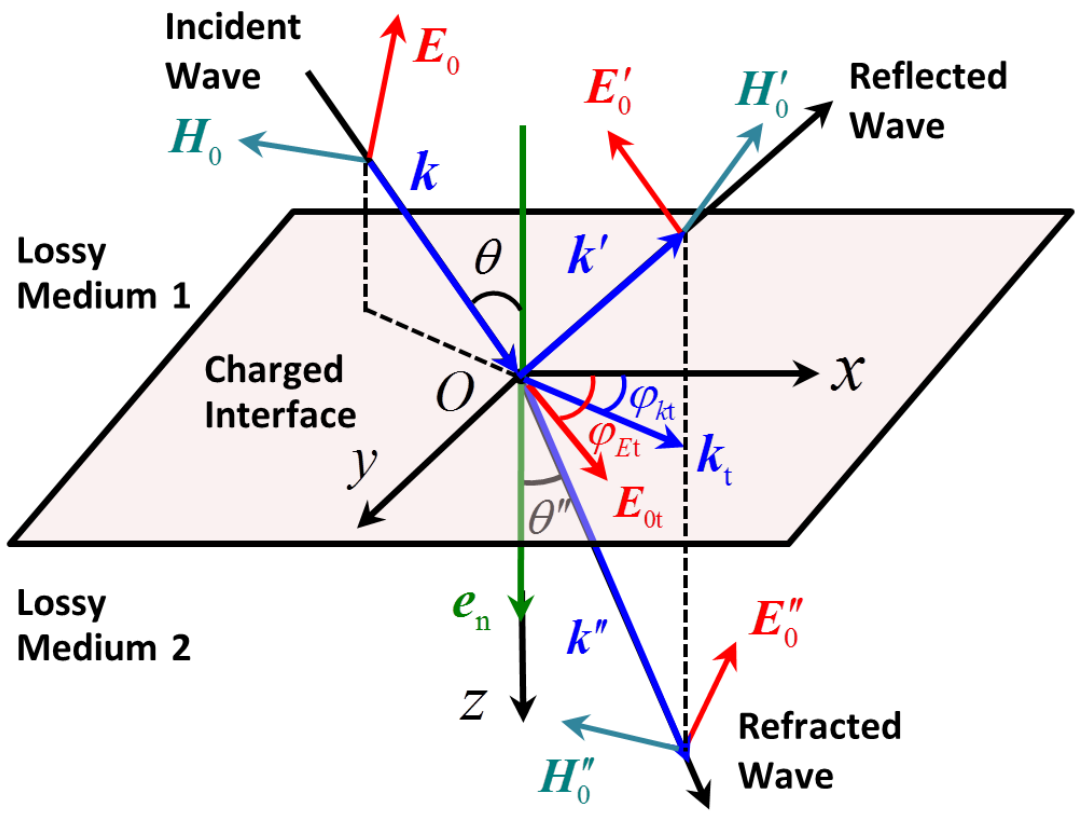

Fig. 3. The incident, reflected and refracted waves are all determined by the tangential electric field magnitude $\boldsymbol{E}_{0\text{t}}$ and the tangential wave vector $\boldsymbol{k}_{\text{t}}$ of an arbitrary plane wave impinges on a charged interface between two isotropic lossy media.

According to (8), (12) and (13), the complex wave vectors of the incident, reflected and refracted waves are

$$\boldsymbol{k} = \boldsymbol{\beta} + \text{j}\boldsymbol{\alpha} = \left(27.2\boldsymbol{e}_x + 15.7\boldsymbol{e}_y + 31.4\boldsymbol{e}_z\right) + \text{j}(28.0\boldsymbol{e}_x + 16.1\boldsymbol{e}_y + 32.3\boldsymbol{e}_z)$$

$$\boldsymbol{k}' = \boldsymbol{\beta}' + \text{j}\boldsymbol{\alpha}' = \left(27.2\boldsymbol{e}_x + 15.7\boldsymbol{e}_y - 31.4\boldsymbol{e}_z\right) + \text{j}(28.0\boldsymbol{e}_x + 16.1\boldsymbol{e}_y - 32.3\boldsymbol{e}_z)$$

$$\boldsymbol{k}'' = \boldsymbol{\beta}'' + \text{j}\boldsymbol{\alpha}'' = \left(27.2\boldsymbol{e}_x + 15.7\boldsymbol{e}_y + 35.3\boldsymbol{e}_z\right) + \text{j}(28.0\boldsymbol{e}_x + 16.1\boldsymbol{e}_y + 31.9\boldsymbol{e}_z)$$

respectively. It can be seen that $\boldsymbol{\beta} \parallel \boldsymbol{\alpha}$ and $\boldsymbol{\beta}' \parallel \boldsymbol{\alpha}'$, so the incident and reflected wave are both homogeneous plane waves. However, $\boldsymbol{\beta}''$ is not parallel to $\boldsymbol{\alpha}''$, so the refracted wave is an inhomogeneous plane wave, which is common for a lossy interface. Based on (22)-(27), the electric field magnitudes of the incident, reflected and refracted waves are

$$\boldsymbol{E}_0 = 70.7\text{e}^{\text{j}1.05}\boldsymbol{e}_x + 70.7\text{e}^{\text{j}1.05}\boldsymbol{e}_y + 96.6\text{e}^{\text{j}-2.09}\boldsymbol{e}_z \text{ V/m}$$
$$\boldsymbol{E}'_0 = 15.6\text{e}^{-\text{j}1.80}\boldsymbol{e}_x + 16.2\text{e}^{-\text{j}1.88}\boldsymbol{e}_y + 21.6\text{e}^{-\text{j}1.83}\boldsymbol{e}_z \text{ V/m}$$
$$\boldsymbol{E}''_0 = 55.9\text{e}^{\text{j}0.967}\boldsymbol{e}_x + 55.0\text{e}^{\text{j}0.985}\boldsymbol{e}_y + 71.9\text{e}^{\text{j}-2.10}\boldsymbol{e}_z \text{ V/m}$$

respectively. Then according to the obtained electric fields and the corresponding magnetic fields calculated by (30), the time-averaged energy flux densities are

$$\boldsymbol{S}_{\text{av}} = 26.8\boldsymbol{e}_x + 15.5\boldsymbol{e}_y + 31.0\boldsymbol{e}_z \text{ W/m}^2$$
$$\boldsymbol{S}'_{\text{av}} = 1.35\boldsymbol{e}_x + 0.780\boldsymbol{e}_y - 1.56\boldsymbol{e}_z \text{ W/m}^2$$
$$\boldsymbol{S}_{\text{av}}^{\text{mix}} = 8.85\boldsymbol{e}_x + 9.49\boldsymbol{e}_y - 1.82\boldsymbol{e}_z \text{ W/m}^2$$
$$\boldsymbol{S}''_{\text{av}} = 23.4\boldsymbol{e}_x + 13.8\boldsymbol{e}_y + 27.6\boldsymbol{e}_z \text{ W/m}^2$$

and the calculated surface Joule heat density at the interface is $p_{\text{s}} = 3.20\times10^{-4}$ W/m$^2$. By substituting these quantities into (31), we can see that the energy balance condition is satisfied and the validity of the proposed formulation is verified. It is also found that the external surface charges have little impact on the reflection and transmission of electromagnetic waves since the surface conductivity is negligibly small with a practical surface charge density.

## V. Conclusion

To summarize, we propose a new formulation for calculating the incident, reflected, and refracted plane waves based on the tangential electric field of the incident plane wave at a charged planar interface between two isotropic lossy media. In contrast to the conventional way based on polarization decomposition, the new methodology decomposes the wave vectors and electric fields of the plane waves into the normal and tangential components with respect to the unit normal unit of interface. The complex wave vectors are easily calculated from the tangential wave vector of the incident plane wave based on the complex angles given by the complex Snell's law. The electric field magnitudes of the incident, reflected and refracted waves are directly derived from the tangential electric field magnitude of the incident plane based on the relationship of the two different types of decomposition and the continuous boundary condition of tangential electric fields. The validity of the proposed formulas is verified by the energy balance condition together with a specific example. We also find that the external surface charges at a charged interface with a practical surface charge density have little impact on the reflection and transmission of electromagnetic waves.

Due to the fact that all time-varying electromagnetic waves can be decomposed into the superposition of time-harmonic plane waves, this work also directly proves the uniqueness theorem of Maxwell's equations and opens a new route for calculating the reflected and transmitted waves at an isotropic, charged and lossy planar interface without the need to perform the polarization decomposition of the incident plane wave of arbitrary polarization state.